\documentclass[english,preprint]{epl}
\usepackage[T1]{fontenc}
\usepackage[latin1]{inputenc}
\usepackage{amsmath}
\usepackage{graphicx}
\usepackage{amssymb}

\makeatletter

\usepackage{float}
\usepackage{amsfonts}
\usepackage{babel}

\makeatother

\title{Is the entropy at the liquid-gas critical point of pure fluids proportional to a master dimensionless constant?}

\author{Yves Garrabos}

\institute{Equipe du Supercritique pour l'Environnement, les Matériaux et l'Espace
- Institut de Chimie de la Matière Condensée de Bordeaux
- Centre National de la Recherche Scientifique
- Université Bordeaux I 
- 87, avenue du Docteur Schweitzer, F 33608 PESSAC Cedex France.}


\date{18 July 2005}

\begin{document}

\maketitle

\begin{abstract}
From a minimal set made of four scale factors defined at the liquid-gas critical point of a pure fluid, and one adjustable parameter which accounts for particle quantum effects, we demonstrate here a master singular behavior of the correlation length for the one-component fluid subclass, using an asymptotic scale dilatation of the physical fields. Such master behavior observed within the preasymtotic domain is in conformity with the renormalized  $\Phi_{d = 3}^{4}$  Field Theory predictions at large correlation length scale of the fluctuating order parameter, for the complete universality class of the symmetrical uniaxial 3D-Ising-like systems. The following consequences are discussed: (i) A comparison between the critical state of pure fluids and the zero-temperature state leads to an intuitive analogy with the (Nerst) third law of thermodynamics, which authorizes specific master form for hyperscaling within the subclass of pure fluids; (ii) A master constant value of the non-dimensional critical entropy can exist for all the pure fluids at the short-ranged lengthscale of the molecular interaction. From this latter hypothesis, we show that the needed four scale factors are the four preferred directions expressing complete thermodynamic (linear) continuity crossing the liquid-gas critical point on the (pressure, volume, temperature) phase surface. 
\end{abstract}

\vspace{-8mm}
The liquid-gas critical point (CP) of a pure fluid is an \emph{unattainable} single point on the $p,v_{\overline{p}},T$ (pressure, particle volume, temperature) phase surface, in which an infinite degrees of freedom are coupled.
As a matter of fact, the critical state at $p_{c}$, $T_{c}$, and $n_{c}=\frac{N_{c}}{V}$ (the critical density number) of a critical amount of matter $N=N_{c}$ filling the volume $V$, is a thermodynamically unstable state, due to the diverging character of the spontaneous fluctuations of extensive variables ($p$ is dual to $V$).
This critical state appears then as a limit of thermodynamic stability at which all the stability determinants (the second derivatives $\frac{\partial^{2}U}{\partial\Omega_{i}\partial\Omega_{j}}$), that were \textsl{strictly} positive for any point of the four-dimensional (4-D) characteristic surface $\phi_{U}\left(U,\Omega_{i}\right)=0$, become zero for CP (see for example \cite{Modell 1983}).
$U\left(\Omega_{i}\right)$ is the total internal energy, while $\Omega_{i}=\left(S,V,N\right)$ are the three associated natural variables. $S$ is the total entropy.

A comprehensive understanding of the diverging character of the spontaneous fluctuations of extensive variables, which characterizes the critical behavior of the one-component fluid close to its unstable critical state, comes from the field theory (FT) framework (see for example \cite{FTbook}).
This theoretical scheme accounts for the infinite degrees of freedom throughout the Hamiltonian of the so-called $\Phi_{d = 3}^{4}(n=1)$-model of the symmetrical uniaxial 3D-Ising-like systems, with associated coupling constant $u_{4}>0$ and finite cutoff wave number $\Lambda_{0}$.
The selected pair of relevant scaling fields, made of the thermal field $t$ weakly fluctuating and the ordering field $h$ more strongly fluctuating, takes exact zero-value at the isolated non-Gaussian (Wilson-Fisher) fixed point $\left\{t = 0; h = 0\right\} $.
At this fixed point, the fluctuations of the (scalar) order-parameter (OP) $m=\left\langle \Phi \right\rangle$, conjugated to $h$, can reach infinite size $\left\{ \xi\sim L = \left(V\right)^{\frac{1}{d}}\sim\infty\right\}$.

This hypothetical situation at $T = T_{c}$ is then only expected for an unphysical pure fluid of infinite size ($V\rightarrow\infty$)
and infinite number of particles ($N\rightarrow\infty$), but with finite number density $n = \frac{N}{V} = \left(v_{\overline{p}}\right)^{-1}$, such that $n_{c} = \left(\frac{N\rightarrow\infty}{V\rightarrow\infty}\right)_{CP}\equiv\frac{1}{v_{\overline{p},c}}$.
Asymptotically close to this critical point, i. e. for small values of $t$ and $h$, power law behaviors of universal features are expected whatever the selected system.
So that, the set of system-dependent parameters which characterizes each one-component fluid is made from i) the critical parameters, such as $T_{c}$, $p_{c}$, and $n_{c}$, ii) the inverse cutoff wave number which characterizes a discrete struture of the fluid particles with spacing $\left(\Lambda_{0}\right)^{-1}$, and iii) the two-scale factors which relate analytically $t$ and $h$ to the respective physical fields proper to each system \cite{Wilson1971}.
 Starting from this asymptotic description of the two-scale universality in the close vicinity of CP, we have postulated in \cite{Garrabos1982}, that the characteristic set of fluid-dependent parameters corresponds to the minimal set of \emph{measured} critical parameters needed to localize CP on the normalized (i.e. 3D) $p,v_{\overline{p}},T$ phase surface. Here normalization refer to particle properties for standard thermodynamics written for a constant amount $N$ of matter.
Our particle notation, i.e. $v_{\overline{p}} = \frac{V}{N}$, uses small letters with explicit subscript $\overline{p}$. 
The minimal set reads $\mathbb{Q}_{c}^{min} = \left\{  T_{c};v_{\overline{p},c}; p_{c}; \gamma_{c}^{'} = \left(\frac{\partial p}{\partial T}\right)_{v_{\overline{p},c}}\right\} _{CP}$ \cite{Garrabos1982,Garrabos1985}.
We can then make dimensionless the thermodynamic and correlation functions of any one-component fluid, using the scale factors, $\left(\beta_{c}\right)^{-1} = k_{B}T_{c}$ for energy unit, and $\alpha_{c} = \left( \frac{k_{B}T_{c}}{p_{c}} \right) ^{\frac{1}{d}}$ for length unit ($k_{B}$ is the Boltzmann constant).
$\alpha_{c}$ is not dependent of the container size $L=\left(V\right)^{\frac{1}{d}}$ \cite{length} and takes a clear physical meaning: $\alpha_{c}$ is the spatial extent of the short-ranged (Lennard-Jones like) molecular interaction.
Therefore, $v_{c,I}=\frac{k_{B}T_{c}}{p_{c}}$, is the microscopic volume of the \emph{critical interaction cell} (CIC).
Introducing the two characteristic dimensionless number, $Z_{c} = \frac{p_{c}v_{\overline{p},c}}{k_{B}T_{c}}$, and $Y_{c} = \gamma_{c}^{'}\frac{p_{c}}{T_{c}}-1$, leads to rewrite the minimal set in the more convenient form, $\mathbb{Q}_{c}^{min} = \left\{ \beta_{c};\alpha_{c};Z_{c};Y_{c}\right\}$.
Now, $\left(Z_{c}\right)^{-1}=n_{c}v_{c,I}$ is the number of particles that fill the CIC. 
$Z_{c}$ and $Y_{c}$ can then take their physical meaning of two-scale factors to formulate the dimensionless master behavior of all the one component fluids asymptotically to their CP \cite{Garrabos1985,Garrabos2002}.

\textbf{"Master" singular behavior of the one-component fluid subclass}

As a matter of fact, asymptotic master singular behavior of dimensionless potentials and dimensionless size of OP fluctuations only occurs when an appropriate scale dilatation method (SDM) is applied to the physical fields, $\Delta\tau^{*} = k_{B}\beta_{c}\left(T-T_{c}\right)$, $\Delta h^{*} = \beta_{c}\left(\mu_{\overline{p}} - \mu_{\overline{p},c}\right)$, and $\Delta m^{*} = \left(\alpha_{c}\right)^{d}\left(n - n_{c}\right)$. $\mu_{\overline{p}}$, dual to $N$, is the chemical potential per particle. In SDM, the ``renormalized'' fields $\mathcal{T}_{qf}^{*}$, $\mathcal{H}_{qf}^{*}$, and $\mathcal{M}_{qf}^{*}$, are proportional to the physical fields, in complete analogy to the FT framework near the Wilson-Fisher fixed point \cite{Wilson1971}. 
For the complete one-component fluid subclass, these renormalized fields are now \cite{Garrabos2005a} defined by
\vspace{-3mm}
\begin{equation}
	\begin{array}{c}
		\mathcal{T}_{qf}^{\ast}\equiv \mathcal{T}^{*} = Y_{c}\Delta\tau^{\ast} \\
		\mathcal{H}_{qf}^{\ast} = \left(\Lambda_{qe}^{\ast}\right)^{2}\mathcal{H}^{\ast} = \left(\Lambda_{qe}^{\ast}\right)^{2}\left(Z_{c}\right)^{-\frac{d}{2}}\Delta h^{\ast} \\
		\mathcal{M}_{qf}^{\ast} = \Lambda_{qe}^{\ast}\mathcal{M}^{*} = \Lambda_{qe}^{\ast}\left(Z_{c}\right)^{\frac{d}{2}}\Delta m^{\ast}
	\end{array}
	\label{eq:field dilatations}
\end{equation}

In Eqs. (1), $\Lambda_{qe}^{*}=1+\lambda_{c}$ accounts for quantum effects on the cut-off parameter for $T\cong T_{c}$ \cite{Garrabos 2005a}.
Writing $\lambda_{c} = \lambda_{q,f}\left(\frac{\Lambda_{T,c}}{\alpha_{c}}\right)$, [$\Lambda_{T,c} = \frac{h_{P}}{\left(2\pi m_{\overline{p}}k_{B}T_{c}\right)^{\frac{1}{2}}}$ is the thermal wavelength at $T = T_{c}$ and $h_{P}$ is the Planck constant], leads to a relative quantum correction of the range of molecular interaction at CP proportional to the ratio $\frac{\Lambda_{T,c}}{\alpha_{c}}$.
Then $\lambda_{q,f}$ appears as an adjustable numerical prefactor which incorporates the quantum particle statistics. 

When critical properties of one standard fluid (xenon \cite{Bagnuls1984}) are known, Eqs. (\ref{eq:field dilatations}) permit to define the constant amplitude values of the master critical behavior \cite{Garrabos1985} in the so-called preasymptotic domain (PAD) \cite{Bagnuls19841987}, where the singular power laws expressed at the first-order of the Wegner expansion \cite{Wegner1972}, are expected to be valid.
Conversely, SDM is then able to estimate all the critical amplitudes appearing in the [two-terms] Wegner expansions for any pure fluid \cite{Bagnuls1984, Garrabos1985}.

\vspace{-4mm}
\begin{figure}[ht]
	\centering
		\includegraphics[width=120.000mm,keepaspectratio]{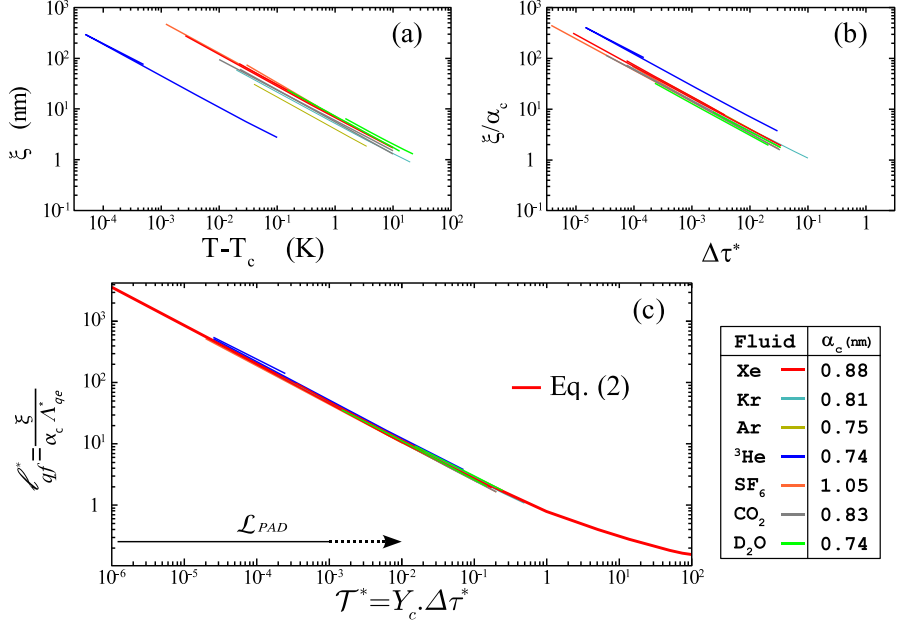}
		\vspace{-4mm}
		\caption{a) Log-Log scale of $\xi^{+}$ (in $nm$) as a function of $T - T_{c} > 0$ (in $K$), along the critical isochore, for $Xe$, $Kr$, $Ar$, $CO_{2}$, $SF_{6}$, $D_{2}O$, and $^{3}He$ (see the respective colors in the inserted Table); b) dimensionless behaviors in units of $\alpha_{c}$ and $\left(\beta_{c}\right)^{-1}$, showing failure of the classical corresponding state scheme; c) master dimensionless behavior of the renormalized correlation length $\ell_{qf}^{*,+}$, as a function of the dilated thermal field $\mathcal{T}^{*}$ [see Eq. (2)]. The arrow indicates the order of magnitude of the expected extension of the preasymptotic domain. Each fluid length scale $\alpha_{c}$ is given in the inserted Table.}
		\label{Fig1}
\end{figure}

\vspace{-2mm}
To demonstrate this important feature where any leading and first confluent amplitudes can be estimated only using $\mathbb{Q}_{c}^{min}$ and $\Lambda_{qe}^{*}$, we are here concerned by the critical behavior of the actual correlation length $\xi^{+}$ along the critical isochore above $T_{c}$ \cite{Garrabos2005b}.
When the \emph{fluid} cutoff wave number $\Lambda_{0}$ is such that $\Lambda_{0}\Lambda_{qe}^{\ast}=\frac{1}{\alpha_{c}}$ \cite{Garrabos2005a}, the renormalized correlation length writes $\ell_{qf}^{\ast} = \Lambda_{0}\xi = \frac{\xi^{\ast}}{\Lambda_{qe}^{\ast}} = \frac{\frac{\xi}{\alpha_{c}}}{\Lambda_{qe}^{\ast}}$.
Within PAD, the two-terms master divergence of $\ell_{qf}^{\ast,+}$ reads as
\begin{equation}
	\begin{array}{c}
 \ell_{qf}^{\ast,+} = \mathcal{Z}_{\ell}^{+}\left(\mathcal{T}^{*}\right)^{-\nu}\left[1 + \mathcal{Z}_{\ell}^{(1),+}\left(\mathcal{T}^{*}\right)^{\Delta}\right]
 	\end{array}
	\label{eq:master correlation length}
\end{equation}
where $\nu = 0.6304\pm0.0013$ and $\Delta = 0.502\pm0.002$ are universal critical exponents \cite{Guida1998}.

The leading amplitude $\mathcal{Z}_{\ell}^{+} = 0.57$ and the first confluent amplitude $\mathcal{Z}_{\ell}^{(1),+} = 0.385$ have constant values for the pure fluid subclass (see \cite{Garrabos2005b} for detailed analysis and the Refs. \cite{Bagnuls1984,Garrabos1985,Garrabos2002} for amplitude values of standard critical xenon).
Here, postulating $\mathcal{Z}_{\ell}^{(1),+}=cte$ for the pure fluid subclass, leads to consider the simpler situation in the $\Phi_{d=3}^{4}(n=1)$-theory, where the starting point for $u_{4}>0$ (in usual renormalized trajectories), is certainly very close to the ideal trajectory between the Gaussian and the Wilson-Fisher fixed points \cite{Bagnuls2000}.

The published fitting results of the correlation length measurements of $Xe$, $Kr$, $Ar$, $CO_{2}$, $SF_{6}$, $D_{2}O$, and $^{3}He$, \cite{ksiexp} have been reported on Figure 1a using dimensioned quantities, which make clearly distinguishable each specific fluid behavior.
Figure 1b gives a representation of the dimensionless quantities obtained from the classical theory of corresponding states, using $\left(\beta_{c}\right)^{-1}$ and $\alpha_{c}$ units. The failure of the classical theory is evidenced from the importance of quantum effects in $^{3}He$ and molecular interaction effects in $D_{2}O$, when comparison to the standard monoatomic $Xe$ is made.
Final representation of the master behavior obtained from SDM is given on Figure 1c.
The scatter between the data corresponds to the estimated precision (10\%) on the determination of each fluid correlation length.

Since $\alpha_{c}$ is a measure of the mean range of interaction forces, Eq. (\ref{eq:master correlation length}) also provides an easy control of the effective extension of the critical domain where the mandatory condition $\ell_{qf}^{\ast,+}\gg1$ is expected to be valid. Then Figure \ref{Fig1}c indicates also the order of magnitude of the PAD length (at least up to $\mathcal{T}^{*}\lessapprox0.01$) where Eq. (\ref{eq:master correlation length}) is valid (see also \cite{Garrabos2002, Garrabos2005b}).
Such result confirms previous similar conclusions based on careful analyses of crossover models and generalized critical fluid e.o.s. \cite{Bagnuls1984,Garrabos1985,Kiselev2003}.

\textbf{Characterization of the thermodynamic "CP vicinity"}

To reach CP from a thermodynamic approach needs to perform an infinite number of transformations between infinite number of near-critical equilibrium states, leading to infinite time to obtain the critical state.
By analogy with the Nernst principle (the so-called third law of thermodynamics), we can also reformulate the previous sentence as follows.
\emph{It is impossible by any procedure, no matter how idealized, to reach exact critical state of any pure fluid in a finite number of operations}.
In a finite number of operations (or finite time), it is then only possible to border the critical point \emph{as
close as possible}, and the final near critical state \emph{at incipient equilibrium} is imposed by the finite sized \emph{container}.
For pure fluids in absence of external field, this finite size of the container is measured by i) finite extensive values of $V$ and $N$, fixing the (small) mean value of $n-n_{c}$ (or $v_{\bar{p}}-v_{\bar{p},c}$, equivalently), proportional to the physical OP; ii) the finite intensive value of $T$, that fixes the (small) mean value of $T-T_{c}$, proportional to the independent physical thermal field.
Therefore, the finite intensive value of $\mu_{\bar{p}}-\mu_{\bar{p},c}$ (or $p-p_{c}$, equivalently), proportional to the independent physical ordering field, is also fixed (from thermodynamics principles).
This finite critical size of the container governs the natural way for the unstable critical fluid to reach incipient stability on an equilibrium state acted by this \emph{near critical container}, the so-called \emph{reservoir} in statistical mechanics.
For such a thermodynamic equilibrium very close to the CP, all the total characteristic potentials $\Pi \equiv \left\{U, H, A, G, J \right\}$ \cite{notation} are homogeneous functions of the first order in terms of their three natural variables among $\Omega_i \equiv \left\{S, V, N, T, p, \mu_{\bar{p}} \right\}$, and Euler's thoerem applies \cite{Modell 1983}.
Correlatively, the Gibbs-Duhem equation, $SdT-Vdp+Nd\mu_{\bar{p}}=0$, requires a normalized description leading to the 3-D representation of thermodynamic equilibrium states. As $N$ and $V$ are two independent extensive variables, in addition to the standard normalization per particle mentionned above, another equivalent normalized scheme occurs using densities
for a system at constant volume. The density notation, i.e. $\pi = \frac{\Pi}{V}$ or $n=\frac{N}{V}$, uses small letters.
 
When the above first-order (equilibrium) scheme is used in the "CP vicinity", the finite critical parameters are basically constitued from the non-zero values of appropriate "first order" derivatives. On this thermodynamic point of view, all the critical parameters, the so-called preferred critical directions in the following, reflect topological continuity of thermodynamics at the CP, directly associated to the proper analytic continuity of the three \emph{intensive} variables $T$, $p$, and $\mu_{\bar{p}}$ when "crossing CP".
For example of our main present concern, the normalized value of 
$s_{\bar{p},c}=\left(\frac{\partial g_{\bar{p}}}{\partial T}\right)_{p,CP}$ [or $s_{c}=\left(\frac{\partial g}{\partial T}\right)_{p,CP}$], should be defined as one among the preferred directions. Obviously, the extended set made with all the critical directions (see below), includes our minimal set hypothetized as containing the needed information to describe critical phenomena (i.e. the needed information to calculate all the leading and first confluent amplitudes of the power law singularities of "second order" derivatives). 
This raises the following question resulting from thermodynamic equivalence between
unattainable critical state and unattainable absolute zero state: \emph{Is
the entropy at the exact liquid-gas critical point of any pure
fluid a characteristic value of each fluid particle, related
to a master dimensionless constant reflecting
universal fluid nature approaching asymptotically CP?} 

Close to the critical point, we are mainly concerned by the fluctuations of $N$ (or $V$, alternatively).
When the energy and the number of particles of a fluid in contact with a particle reservoir can fluctuate, the basic link between statistical mechanics and thermodynamics of a fixed volume of matter, where $T$ and $\mu_{\bar{p}}$ are the operating variables can be correctly established only from the Grand canonical statistical distribution.
The Grand potential $J\left(T,V,\mu_{\bar{p}}\right)=-p\, V$ is then naturally selected to characterize the equilibrium state of
the system maintained at constant volume. Its equilibrium state is characterized by the Grand potential density $j\left(T,\mu_{\bar{p}}\right) = \frac{J}{V}$, whose opposite, i.e. $p\left(T,\mu_{\bar{p}}\right)$, gives the characteristic surface close to the CP schematized in figure 2 \cite{Gibbs}.
This result complements the Canonical statistical description connected to the Helmholtz free energy $A\left(T,V,N\right)$, where the Helmholtz free energy per particle $a_{\overline{p}}\left(T,v_{\overline{p}}\right)= \frac{A}{N}$ characterizes the equilibrium state of the system of constant amount of matter in contact with an energy reservoir (a thermostat).
To discuss differences between minimal and extended sets, needs to consider the e.o.s. pairs, $\left\{ p\left(T,v_{\overline{p}}\right); s_{p}\left(T,v_{\overline{p}}\right)\right\} $, associated to $a_{\overline{p}}$, or, $\left\{ n\left(T,\mu_{\overline{p}}\right); s\left(T,\mu_{\overline{p}}\right)\right\} $, associated to $j\left(T,\mu_{\overline{p}}\right)$.
However, only their common $\mu_{\overline{p}};T$ diagram contains three new unmeasured characteristic parameters, in addition to the measured ones in the usual $p;T$ diagram (see the respective binary diagrams constructed in Figure 2):

 i) $\mu_{\overline{p},c}^{*}\equiv g_{\overline{p},c}^{\ast} = \beta_{c}\mu_{\overline{p},c}$, with $a_{\overline{p},c} = j_{\overline{p},c} + g_{\overline{p},c} = -p_{c}v_{\overline{p},c} + \mu_{\overline{p},c}$ and $a_{\overline{p},c}^{\ast} = j_{\overline{p},c}^{*} + g_{\overline{p},c}^{\ast} = -Z_{c}+\mu_{\overline{p},c}^{\ast}$; $\mu_{\overline{p},c}$ is another energy scale factor (in addition to $\left(\beta_{c}\right)^{-1}$);
 
 ii) $x_{\overline{p},c}^{\ast}=\frac{\delta_{c}^{'}}{k_{B}}$ (obtained without use of $\left(\beta_{c}\right)^{-1}$ and $\alpha_{c}$),
where the preferred direction, $\delta_{c}^{'}=\left[\left(\frac{\partial g_{\overline{p}}}{\partial T}\right)_{v_{\overline{p}}}\right]_{CP}$,
characterizes thermodynamic continuity in this diagram ;

 iii) $s_{\overline{p},c}^{\ast}=\frac{s_{\overline{p},c}}{k_{B}}$ (also obtained without use of $\left(\beta_{c}\right)^{-1}$ and $\alpha_{c}$),
where the critical entropy per particle $s_{\overline{p},c} = -\left[\left(\frac{\partial\mu_{\overline{p}}}{\partial T}\right)_{p}\right]_{CP}>0$
is a preferred direction connected to $\gamma_{c}^{'}$ (measured on the $p;T$ diagram, Fig. 2a), and $\delta_{c}^{'}$ (unmeasured on the $\mu_{\overline{p}};T$ diagram, Fig. 2b), throughout
the relation $s_{\overline{p},c}=\gamma_{c}^{'}v_{\overline{p},c}-\delta_{c}^{'}$, leading to  $s_{\overline{p},c}^{\ast} + x_{\overline{p},c}^{\ast} = Y_{c}Z_{c}-j_{\overline{p},c}^{\ast}$, with $-j_{\overline{p},c}^{\ast}=Z_{c}$.

The product $Y_{c}Z_{c} = \frac{\left[v\left(\frac{\partial s_{\overline{p}}}{\partial v}\right)_{h = h_{c}}\right]_{CP}}{k_{B}}$ is independent of the reduction process, then is a particle property, characteristic of each one-component fluid.
De facto, the non-dimensional pair $\left\{ Z_{c},Y_{c}Z_{c}\right\} $ (obtained without use of $\left(\beta_{c}\right)^{-1}$ and $\alpha_{c}$) is associated to the two preferred directions of $\Phi_{j_{\bar{p},V = cte}}\left(j_{\bar{p},V = cte},T,v_{\overline{p}}\right) = 0$, where $j_{\bar{p},V = cte}\left(T,v_{\overline{p}}\right) = \left(\frac{J}{N}\right)_{V = cte}$ is the Grand potential per particle of a fluid maintained at constant volume.

Finally, the extended set $\left\{ T_{c}; p_{c}; Z_{c}; Y_{c}; \mu_{\overline{p},c}^{\ast}; s_{\overline{p},c}^{\ast}\right\} _{CP}$ is the complete set from which we are able to calculate all the other
characteristic critical parameters using linearized thermodynamics.
For example, from $U=TS+G+J$, we obtain $\frac{u_{c,\overline{p}}}{T_{c}} = s_{c,\overline{p}} + \frac{\mu_{c\overline{,p}}}{T_{c}} + \frac{-p_{c}v_{0,c}}{T_{c}}$ and then $u_{c,\overline{p}}^{\ast} = \mu_{c,\overline{p}}^{\ast}+s_{c,\overline{p}}^{\ast}-Z_{c}$. 

\begin{figure}[ht]
	\centering
		\includegraphics[width=120.000mm,keepaspectratio]{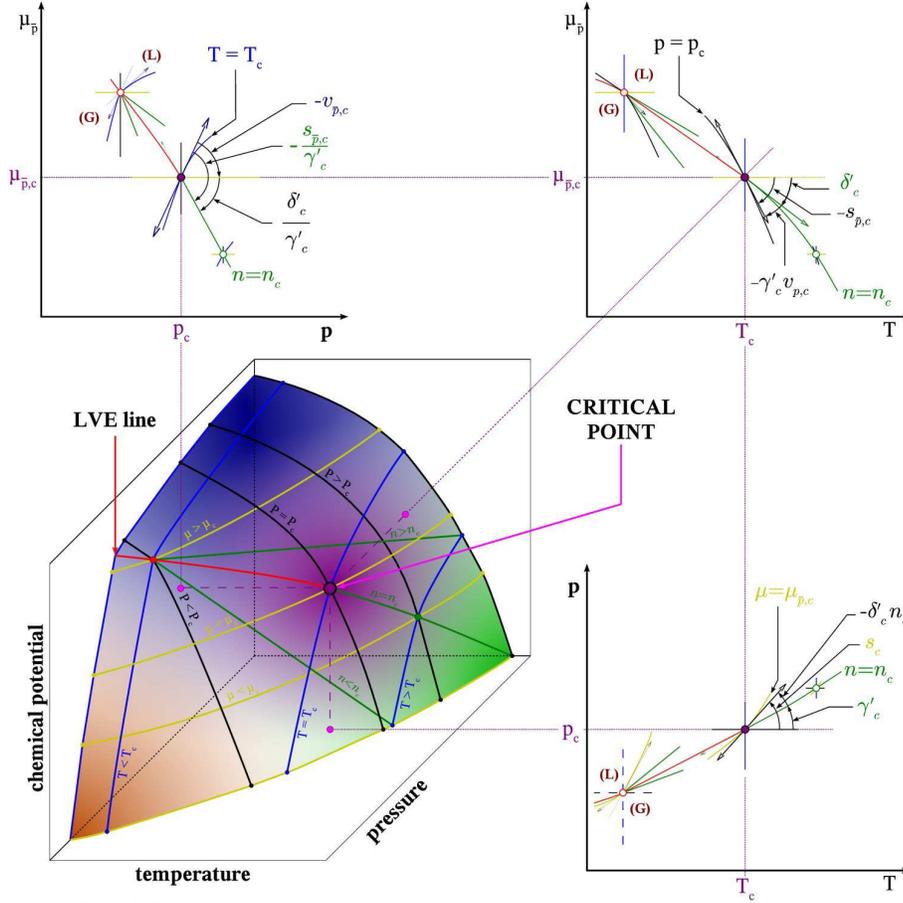}
		\vspace{-5mm}
	\caption{Preferred directions crossing critical point in $\mu_{\overline{p}};p$ (a), $\mu_{\overline{p}};T$ (b), and $p;T$ (c) diagrams obtained from projections of the 3D characteristic surface $p\left(T,\mu_{\bar{p}}\right)$and corresponding isothermal (blue), isobaric (black), isochoric (green), and isochemical potential (yellow) isoclines. Non-homogenous two-phase domain corresponds to the liquid-vapor equilibrium (LVE) (red) line. See also \cite{Gibbs}.}
	\label{Fig2}
\end{figure}
\vspace{-2mm}

We can now reformulate critical thermodynamics, at the scale of the microscopic interaction cell of volume $v_{c,I}=\frac{k_{B}T_{c}}{p_{c}}$ filled with $\frac{1}{Z_{c}}$ particles, where we expect that \emph{all the system information is contained}. Multiplying then the above particle relations by $\frac{1}{Z_{c}}$, we obtain
\vspace{-3mm}
\begin{equation}
	\begin{array}{rcl}
		j_{c,I}^{\ast} & = & -1 \\
		s_{c,I}^{\ast}-1 & = & Y_{c}-X_{c} \\
		u_{c,I}^{\ast} & = & s_{c,I}^{\ast} + \mu_{c,I}^{\ast} - 1
	\end{array}
	\label{eq:master s}
	\vspace{-3mm}
\end{equation}
where $X_c = \frac{x_{c,\overline{p}}^{\ast}}{Z_c}$.
As expected, $Z_{c}$ disappears in the above equations. $X_{c}$, as well as $Y_{c}$, are two characteristic properties per CIC. 

Since the Grand potential (which favorizes the local ordering) takes master value at criticality, $j_{c,I}^{\ast}=-1$, whatever the selected CIC volume, it seems natural to postulate that the entropy at criticality (which favorizes the local disorder) can also take a master constant value, $s_{c,I}^{\ast}=const$.
From Eqs. (\ref{eq:master s}), $\mu_{c,I}^{\ast}-u_{c,I}^{\ast}$ is then also constant.
More generally, all the remaining critical free energies per CIC volume, $u_{c,I}^{\ast}$, $g_{c,I}^{\ast} \equiv \mu_{c,I}^{\ast}$, $h_{c,I}^{\ast} = u_{c,I}^{\ast}-j_{c,I}^{\ast} = 1 + u_{c,I}^{\ast}$, and $a_{c,I}^{\ast} = j_{c,I}^{\ast} + g_{c,I}^{\ast} = -1 + \mu_{c,I}^{\ast}$, are also master constants, except a possible common constant value corresponding to an energy translation. Finally, the two dimensionless preferred directions $X_{c}$ and $Y_{c}$ are related. The remaining set of the characteristic parameters which contains the complete information at the CIC scale is well  $\mathbb{Q}_{c}^{min}$. 
Our above suggestion complements the third law of thermodynamics ($s_{\overline{p}} = 0$ at $T = 0$), and defines particle entropy and particle free energy from (unknown) constants of proportionality to $Z_{c}$ at exact CP, in conformity with basic thermodynamic principes. Therefore $\left(\beta_{c}\right)^{-1}$ is the unique fluid-dependent scale factor for energy, as $\left(\alpha_{c}\right)^{-1}$ is the unique fluid-dependent scale factor for length, as initially postulated in \cite{Garrabos1982}.

As a conclusion, from a formulation of \emph{critical thermodynamics} in units of the properties of the CIC volume, the above analysis shows that the appropriate scale dilatation of the physical fields seems adequate to observe scaling of their asymptotic critical singularities (with only one adjustable parameter to account for quantum effects). To describe the fluid singular behavior within the preasymptotic domain, scale dilatations of two independent fields can then be used as controlled simplifications of linear combinations of three fields \cite{Kim2003} associated to the density formulation of non-symmetrized thermodynamic potentials at finite distance to the CP.


\begin{thebibliography}{19}
\bibitem{length}This is not the case of the second critical length scale defined by $\frac{m_{\overline{p},c}}{rho_{c}}$.
\end{thebibliography}
\end{document}